\documentstyle[12pt]{article}
\title{Unmissing the Missing Matter\\ in Neo-Kaluza-Klein
Universes\thanks{A preliminary version}}
\author{ Hai 
Halevi\thanks{c/o G. K\"albermann, Hebrew University, Israel; e-mail:
hope@vms.huji.ac.il}}
\begin{document}
\maketitle

\begin{abstract}
Uncompactified KK universes are so intrinsically connected to the
otherwise only empirically required ``missing" Dark Matter (DM),
that:

\noindent 1) They yield a simple prediction which explains both
the enigma of the extra-dimensions' (XD) unobservability and the
enigma of the present DM. The two enigmas are ``annihilated" into
the hypothesis of ``missing light", or better of ``photonland".
This eliminates the very need to hypothesize/search/find
DM-candidates of exceptional/exotic properties to explain their
darkness.

\noindent 2) An early, spontaneous gravitational XD-collapse of their
natural 5D-DM replaces KK's compactification mechanisms and
cylindricity condition, and (partly?) eliminates the quandary of
the radically insufficient density fluctuations.
\end{abstract}
\noindent {\bf (I)} Recently there has been renewed interest and
progress in the KK-5D-cosmology. In many of the new variations:

\noindent A) Artificial, ad hoc geometrical preconditions are
eliminated, to be eventually replaced by physical processes.
Thus, unless otherwise proved

\noindent B) The present universe -- not merely its primordial
``ape"-- is occupied by 5D-bodies: KK-solitons (5D blackhole
analogues) \cite{wes94} and/or others [e.g. 2], if not even $(n>5)$
D-densities.

Since these two changes enable fundamental advancements and
research directions, we label these universes ``neo-KKU".  Among
other things, one can see that:

\noindent 1) 4D-GR is {\bf embedded} in a flat 5D-space
\cite{wes942}, thus curvature can also be seen as exterior.

\noindent 2) There exists an equational equivalence between empty
5D-GR and 4D-GR with sources \cite{wes92,ovw97,mcm94}. This
stresses the necessity to decipher {\bf how} could our universe
have appeared from pure geometry [e.g. 7] (Einstein's famous
vision). In the old-KK paradigm geometrical conditions helped the
vision to resemble a miracle. Neo-KK workers have invested much
efforts [e.g. 5,8,9] in order to uncover the {\bf physical}
processes behind this ``induced matter'' interpretation. However,
many still hope this ``induced" matter to be ``a manifestation of
pure geometry" \cite{ovw97}.

\noindent 3) As the mathematics of ``inducing"
4D-GR plus sources from empty 5D-GR was generalized
\cite{chs93,rrt95,cmm95} to nD, the neo-KK paradigm automatically
includes universes which generate/allow dimensional collapses from
$(n>5)$D; and into $(n<4)$D densities/fluctuations/particles.
This allows condition B above, and our hypothesis below, to assume
non-zero densities of $n > 5$.

\noindent {\bf (II)} Independent of the KK paradigm, cosmological-theoretical {\bf
and} astronomical-observational discrepancies almost compel us to
theorize and find DM-candidates:

\noindent a) Calculations of galaxy/cluster sizes definitely
require a prebaryonic collapse (but see more below). Since  the
source of such a collapse is an enigma, theories were
artificially/ad-hocly stretched to hypothesize the existence of
prebaryonic matter.

\noindent b) The presently known matter makes up only a small
percentage of the gravitation needed {\bf today} to prevent the
galaxies from escaping their clusters, and from losing their own
peripheries.

\noindent c) It was recently shown that ``any attempt to explain
the mass discrepancy\dots using alternative gravity instead of DM
comes at the price of having to abandon the equivalence principle.
DM can know be seen in a new light: it is indispensable\dots"
\cite{ede99}.

Attempting to replace the ad-hocly tailored geometrical
transformations from 5D to 4D-GR, by physical processes by which
5D-densities collapse/confine (see below) to 4D, we can no longer
appriorily expect them to be elegantly perfect and total. As
condition A generates condition B -- unless proved/assumed to
have been completely collapsed into 4D-densities --- it is {\bf
essential} to analyze and conceive our present universe
five-dimensionally. This is of great importance in two major
respects:

\noindent 1) In old-KKU the primordial 5D-densities are
almost\footnote{Compactified 5D-KK radically $(10^{22})$ disagree
\cite{ovw97} with observations, but see below.} untestable:
geometrical preconditions ad-hocly {\bf equate} them to GR's
4D-densities, which denies us
differential-hence-testable-predictions. By contrast, in neo-KKU
the possibility (shown to be a necessity) of great {\bf present}
abundance of 5D-densities greatly differs from any conventional
prediction, and is shown to be testable.

\noindent 2) These present-universe-5D-densities are shown to be
{\bf natural} DM-candidates; are extensively investigated as
solitons [e.g. 1, 7, 14, 15] and other forms of ``extended
objects"; are proved below to yield more than DM-candidates.

The neo-KK's development is still partial.  Firstly, only {\bf
indications} for such physical processes were found. Secondly,
there still exist reservations/questions/ unclearities concerning
our validity in analyzing physics 5-dimensionally
\cite{wes942,ovw97,bil96,wel95} (but see below). Thirdly ---
without entering this complex issue here --- the appearance of our
physical universe from {\bf absolutely}\footnote{In \cite{halev}
we try to extend it to 5D {\bf fluctuations} of $(\pm)$
curvatures/densities {\bf around} zero.} empty 5D-GR might remain
a mathematical ``miraculous\footnote{``The fact that\dots the
sourceless field equations lead to \dots {\bf with} source matter,
constitutes the central miracle of Kaluza-Klein theory."
\cite{ovw97}} creation" of 4D-``something" from 5D-``nothing".

However, one thing is clear: if a 5D-universe (5DU) does not merely ``induce" or
``manifest as" \cite{ovw97}, but physically {\bf produce} our
4DU -- these collapse processes (see below) also take time.  If they started
before our 4DU has emerged, neo-KKU intrinsically predicts extra
prebaryonic {\bf time}. This prediction neither contradicts nor
necessitates extra prebaryonic {\bf matter}. Therefore, unless
compelled by a -- c  above, some versions of neo-KKU could
predict {\bf no} prebaryonic missing matter.\footnote{Even more so
in time-dependent-G-models, and other models which allow/produce
extended inflations.} Secondly, as the prebaryonic collapse ---
{\bf and} its smaller galaxy horizons \cite{wes92} --- are
(presently) integral parts of neo-KKU, they seem to break down if
the prebaryonic collapse is refuted. Surprisingly, however, a
several hundred thousands year long gravitational XD-collapse is
predicted below. If verified, there {\bf might} be no prebaryonic
matter {\bf and} no prebaryonic collapse --- only prebaryonic
time, in which only inflationary expansion takes place, {\bf without}
``mechanisms \dots to pump entropy into the 4DU" \cite{ovw97}.

On the other hand, the {\bf present} universe seems to contain
vast amounts of DM. While several approaches exist in
accepting/rejecting various candidates, {\bf everybody} (still)
believes that their unobservability is due to some special or even
exotic properties. This has driven us to theorize DM-candidates
which are not intrinsically predicted (and sometimes even not
tolerated) by physical theories, unless stretched/burdened by
extra- theoretical matters$\dots$ In sharp contrast, in neo-KKU
the primordial matter needs only remain in its natural 5D-density
to form the present-epoch-DM!  So, as we are compelled to predict
both the prebaryonic collapse and the great present abundance of
5D extra matter, just a simple variation is needed: The
prebaryonic collapse from 5D to 4D (or from $n\geq 5$ to $n\leq4$,
but see below) was {\bf partial} --- all the 5D density was
condensed, yet only a small part of it fully collapsed to (or
arrested in, below) 4D.

A partial collapse might at first look strange. Yet, ``who
ordered" a perfect ``production line" which does not cease before
{\bf all} its 5D ``raw material" became fully collapsed to {\bf
our} 4D ``specification"?! (see also \cite{wes942,mcm94}). Only
the inflationary expansion/XD-collapse processes dictate the
collapse ratio of 4D-matter/5D-DM. So, by contrast to old-KKU,
{\bf all} neo-KKU must obey (and eventually  predict) the
DM/matter ratio! In addition, this also upgrades the paradigm's
refutability: Any deviation from empirical constraints on the
present DM/matter ratio {\bf refutes} the neo-KK paradigm.  While
these constraints are still vague, satellites new observations are
expected to rapidly clarify them.

Predicting the present
DM-ratio -- instead of only explaining the DM phenomena -- will
upgrade the KK-paradigm to eventually become fully testable. But
prior to this, we must realize that our predicted, natural {\bf
neo-KK's nD-DM are presently bugged by two main drawbacks:}

\noindent 1) Explaining/predicting the missing-dark-matter by
missing-dark-dimensions does not contribute too much conceptually
--- unless the XD-unobservability itself is predicted. Yet, old-KKU
rely on ad-hoc explanations of it.  E.g. [5] ``They are a priori cylindrical.
No mechanism is suggested to" (sufficiently) ``explain why physics
depends on the first four coordinates, but not on the extra ones".

\noindent 2) Hypothesizing a natural 5D-DM, we ascribe ``5D
feathers" not only to the primordial universe, but to our present
4DU.  This is a bizarre prediction; much more than the elegantly
structured solitons show. If such ``extended-feathers" have some
90\% of the total messenergy, our 4DU resembles a small
4D-body of a vast, surrealistic peacock. But beyond being bizarre,
this prediction is dangerous.  As 5D-densities vary by $1/R^4$
\cite{wes94} --- and nD-densities by $1/R^n!$ --- nD
``unobservable peacocks" might be of monstrous messenergies, no
less than ``our regular" 4D-blackholes.  Therefore, by ``just"
proposing $n>5$, these elegantly hidden nD-blackhole-analogues could
have ``saved" us even if we had to explain 99.99\% of DM!

Reflecting only the {\bf assumed} XD-unobservability, and
dangerous in higher D models, KK's nD-DM cannot be trusted before
we are able to {\bf predict} the XD-unobservability; and {\bf here
our main hypothesis begins.}

\noindent{\bf (III)} The present work proposes a more essential connection
between XD and DM.  It deciphers and eliminates the enigmas of both the DM
and the unobservability of XD(s).  It is tested against the CMBR's photon
density, and drastically reduces the amazing present insufficiency of the
primordial density-fluctuations.  And above all, it clarifies that it is
the behavior of our friends the photons -- not of new/exotic candidates --
which restricts our observations to 4DU!  Thus all XD-matter (XDM) "becomes"
(simply remains) totally unobservable.

As 5D-solitons/other  natural nD-DM are (trivially) predicted and no artificial
pre-condition "irons" the XD, their width in XD might be large or small, but not
negligible as in the old KKU.   This (alone) yields a simple fundamental possibility
hence hypothesis: If {\bf photons}\footnote{But not the massive {\bf virtual} photons.}
are intrinsically {\bf unable} to propagate in XD (s), a
great fraction of any neo -KKU matter cannot exchange photons with us, which makes
us {\bf XD-blind!}  Since XDM does not even absorb photons, it cannot distort our
bolometrical readings, except by lensing. Thus, XDM is not just dark (like
baryonic DM-candidates) but bolometrically {\bf traceless}.

 While in old-KKU gravitation in XD (artificially) loses its significance, in neo-KK it
 does not. By contrast (to gravitons ?!), photons are here hypothesized -- and (almost)
 empirically proved below -- to have zero XD-velocity.  If XDM cannot exchange photons
 with us, but does interact gravitationally with us, the (great) discrepancy which dictates
 the "missing matter" is predicted!

 Our hypothesis is in essential harmony with Wesson \& Liu's hypothesis [18] (below)
 that massive particles' electric charge is (a function of) their XD- momentum.  
 It is however in
 clear contradiction with Gogberashvili's new 5D-theory [19] which seems to rely on
 photons' XD-velocity.  As the neo-KKU are quite testable [e.g. 4,7,14,20-22] much of
 this disagreement can be resolved empirically (see below).

 To prove our hypothesis we shall use the radiation and matter-dominated-eras as our lab.
 If photons had velocity in XD, their (strong) radiation-pressure in the radiation
 dominated era would have hyper-scattered the entire fireball in the 5D 
 hyperspace\footnote{Although one may assume the XD to be smaller 
 than the relevant mean free paths,
 this in itself cannot prevent any radiation-matter interaction. }.
 As its density would then diminish by a higher {\bf power}\footnote{See 
 for example 5D-solitons' density  $\propto  1/R^4$ [1].}
 of their radius, the fireball's
 (radiation + matter) densities and temperatures are definitely predicted to drop much
 faster than by any standard cosmology.  This drastically shortens "our"   300,000 years'
 radiation dominated period, and recombination takes place.  As photons decouple from
 matter, their expansion is no longer decelerated by the matter densities.  This
 transparency is crucial to our hypothesis:  it turns the photonball absolutely
 free to hyperexpand in nD.  So, {\bf if} photons {\bf had} XD-velocity, they would have been
 at least 5D-hyperscattered, decreasing their density and temperature significantly
 below the standard predictions.  This would definitely contradict the
 (already successfully predicted) observed photon's number-density and
 temperature in the CMBR!\footnote{Beyond changing the particles mean free paths 
 values and relations to the altered radia of the universe, thereby disrupting (quite)
 established and accepted cosmological phases, hence also the universe's age.}

Since the above cosmological eras are not "ideal labs", it is unclear whether their
radiation's and matter's expansions in XD's are cut off by the photons'/field's own
properties or by some external forces.  We do not attempt (here) to specify the first,
but to refute the second. At the radiation dominated era the prebaryonic collapse
processes are definitely over, thus leave the photons free to expand in XD.  But, one
may cautiously suspect the yet insufficiently known inflationary XD-collapse (but see
below) to have extended into the radiation dominated era.  However, any such pressure
is related to/ represented by a non-zero cosmological constant.  At this era -- and
certainly at the far later transparent era which is crucial here -- the cosmological
constant vanishes, hence nothing counterforces the radiation pressure.  While several
possible $\Lambda$-calculations exist -- it is safe to assume that at this 300,000
year long era the $\Lambda \sim 0$ related field(s) become weaker than the matter fields,
which are themselves weaker than the radiation pressure. To quote from a recent
10D-KK example, `` At the end of the inflationary phase transition, a radiation
dominated era is brought about by demanding the total effective cosmological
constant to vanish" [23].

With (almost?) no external pressures in XD, only the field's own properties can cut
off the photons' XD-velocity! So, our best friends the photons are self-confined to 4DU.  
{\bf Our} photon-dependent
{\bf observations} are therefore encapsulated in ``photonland"!  ``Luckily", photonland
is not flatland, but overlaps our 4DU.

 If photons were merely oppositely charged pairs, Wesson
\& Liu's hypothesis would suffice to predict their annihilations to cancel their
XD-momenta, hence XD-velocities (see below).  While this is more complicated
(and may involve 5D-field processes) photons (at least) behave as if they were
oppositely charged pairs, which cancel each other's charge and opposite 
XD-momenta\footnote{Photons might vibrate-but-not-propagate in XD, 
{\bf as} they do in the ordinary two dimensions orthogonal to their propagation.}
 -hence self confined to 4D.

Beyond being unobservable, the quantum processes which produce the photons' zero velocity
and radiation pressure in XD, are presently hidden "underneath" the elegant-looking,
schematic condition XD=constant [7].  This exemplifies the fundamental importance of neo-KK
approach to try to replace even simple and reasonable schematic preconditions by physical
processes.  However, {\bf when} oppositely charged particles annihilate into photons/other
electrically natural particles (ENP), Wesson and Liu's hypothesis [18] suffices.
But, most-if-not-all the fireball's photons' were produced this way.  Whether at the
hadron era's equilibrium reaction ($\gamma + \gamma \leftrightarrow$ lepton +
antilepton); at the lepton era's muons' disappearance ($\mu_+ \mu_- \to 2\gamma$); or even at the vast/total positrons'
disappearance ($e^+ + e^- \to 2\gamma$) by the excess of electrons (presumably) produced
by the very early anti-proton decay.  If valid, this 5D-process takes part in the universe's earliest
(KK) dimensional collapse: from massive and charged particles with {\bf 5D}-velocity
(``5D-plasma''), to photons and other ENP with velocity
in only 4D.   This transforms the KK paradigm, from
the mathematically possible\footnote{The mathematical equivalence [4,6] alone does not
demand any creation.  As stressed by Wesson [e.g 3] even  the 4D appearance might not
mean creation, but rather how 4DU/5DU appears to 5DU/4DU-observers.}
``induced (4D) matter from empty 5D-GR",
to the physically predicted ``produced (4D) ENP from ($\pm$)charged 5D-densities". The
``5D-plasma'' may require 6D-empty-GR (see I, 3) or 5D ($\pm$) curvatures /fluctuations around
zero [17]. This transformation will take us even further in chapter (V).

Notice that the fireball's distribution is not predicted to be 4D.  Only the ENP's 
XD-velocity
is paralyzed.  This is why a  neo-KKU {\bf remains} 5D, yet bolometrically disconnected:
photons do not illuminate all matter.  By contrast -- as gravitation is not
XD-paralyzed, the XD-collapse processes continue, thereby narrowing the XD-distribution
(but much more in IV).

Notice also that nD-KKU analysis of earlier eras might easily refute the entire neo-KK
approach -- or at least impose constraints on n -- especially if found to disagree with the observed (and successfully predicted) cosmic abundance.  The neo-KK approach hence owes to conduct this complicated examination regardless of our specific photonland hypothesis.  Since we do not predict the virtual
photons to have zero-XD velocity, such an examination will not help here
(see also [24]).

\noindent {\bf (IV)}
 The photonland hypothesis is intrinsically {\bf compelled to predict} another
XD-physical-process.  Namely, if photons have no velocity in XD, a "radiation dominance"
can never exist in XD(s)!  Lacking radiation pressure in XD, even the hottest photonball
cannot stop the matter's gravitational collapse along XD(s), as it does in 4D.  Thus
``photonland" definitely predicts that the 5D-fireball's matter has {\bf already} collapsed
in XD since its very appearance -- "preparing ground" (mainly ENP densities, see below)
for the galaxies'/clusters' formation -- along the {\bf entire} radiation dominant era.

While this self-imposed, strange prediction of an ``extended collapse" might easily
complicate/refute the photonland hypothesis -- it might no less than replace old-KKU's
artificial/hardly physical compactification mechanisms! Unlike the "cylindricity"
abstraction, which is either unphysical, or hyperenergetic to an extent that generates
radical disagreements with observations [e.g. 7], this natural gravitational collapse
process neither requires a circular typology nor is required to ``squeeze out" the XD(s)
for us!  With a mass-E-distribution along a not-at-all-negligible XD, our 4D-hypersurface
is gradually produced -- just as the galaxies/clusters are formed when - much later -
photonland is freed from photon-dominance!

But moreover, (without trying to prematurely\footnote{These early XD(s) collapse 
processes might take various forms, and even invoke Alfvenic components.} 
develop a detailed picture) this 300,000
year period of undisturbed XD-collapse obviously mitigates -- if not eliminates! --
the harsh quandary of primordial density-fluctuations, which are presently {\bf radically}
insufficient for galaxy formation.   A variety of KK inflationary
scenaria is now expected to eliminate the insufficiency trouble; and a 5D one might
suffice [17]. {\bf And},  while this extreme insufficiency has hitherto
compelled us to predict only a prebaryonic collapse - ``photonland" (partially/totally)
replaces it by it's long pretransparent\footnote{Thus, the pretransparent collapse, 
not the prebaryonic 
one, must eventually predict - and be tested   against - the above DM/matter ratio.} 
DM-collapse!  
If [18] is valid, this pretransparent gravitational formation of our
4D-bubble is partially
a dimensional collapse, since the 5D-free charged-pairs annihilate into the XD-paralyzed
photons/ENP. This is crucial in determining whether/when the
hypersurface's XD-thickness (and our XD-blindness) ceases to be
macroscopic, and remains significant ``only'' to quantum processes
[24].

Finding "photonland" predicting/affecting/speculating fundamental cosmological processes, 
we must try to refute /prove it in all possible ways; especially by:

\noindent 1) Quantifying at least the ($e^+ + e^- \to 2\gamma$) 5D-processes, 
and their resulting densities, pressures, temperatures, and mean free paths' 
relationships to the universes' radia.

\noindent 2) Quantifying our radically extended pretransparent XD-collapse processes, and
their variety of possible disagreements with standard cosmology, the present DM/matter
ratio, and particle physics. Notice that if the dimensional
collapse is shown {\bf not} to be partial-this is a full
compactification. ``Photoland'' ceases to explain the DM, and
reduces to a compactification mechanism candidate.

\noindent 3) Upgrading Wesson \& Liu's hypothesis to explain (and eventually predict) how the
massless and chargeless photons must have zero XD-propagation, while the massless
and chargeless gravitons do have XD-propagation, hence are clearly insensitive to XD.
Please recall that gravitons are as admissible in neo-KKU as photons are [e.g. 5].

The photoland hypothesis is clearly contradictory to Gogberashvili's new theory  [19],
which (apparently?) relies on photons' non-zero velocity in five dimensions.  He must
therefore explain/predict (see above) how his theory does not destroy the conventional,
successful predictions of both the radiation and the matter dominated eras'
processes/abundance, and the observed density and temperature of the present CMBR.

\noindent {\bf (V)} Beyond predicting the gravitational-bolometric XD-gap, hence unmissing
the (presently called) ``missing matter" {\bf and} the vast primordial densities-photonland 
is also our {\bf observational prison} which is of an (annoying) fundamental
importance to physics.  In neo-KKU, 4D-GR is embedded in a flat hyperspace; its'
4D-densities interact with the 5D-solitons/ other ``extended objects"; but our
photon-dependent-observations are confined to 4DU.  {\bf This} is not merely ``an
unfortunate choice of coordinates" [7].  If the photonland hypothesis is
valid, quantum processes wipe out our illusion\footnote{This inversely resembles 
the quantum processes which wipe out the illusion of certainty in observations: 
As photons cannot push-hence-distort our XD-observational-targets, we cannot observe them.}
 of choosing observations
beyond the photons' geometrical self-confinement!   While this involves
deeper philosophy of science which seems inappropriate here-choosing to
add ``photonless observations" of XDM  thereby developing a higher dimensional perspective,
is not unscientific/unreliable just because it transcends photonland.  On the contrary,
our photon-dependent-perspective is intrinsically biased; this has caused physics
to miss most of the universe's matter, and to misperceive our 5DU as a narrow 4DU.
If/as 5D-DM interacts with us (at least) gravitationally, XDM and 5D-densities are
neither unreal, nor untrustable. Even under the extreme assumption that all real
particles are 4D-bodies/fluctuations, and all 5D-DM are only virtual particles
[24], their systematic analysis in a 5D frame of reference is ``meta-optica", not
``metaphysica".  Hence, if our {\bf present} universe is 5D, the possibilities of
5D massive/massless, charged/chargless, virtual particles (if not even quarks
and leptons) must be vehemently scrutinized.  Being XD-blind -- unmissing the missing 
matter must also imply undismissing most of our universe's matter and volume as merely 
``dark"!

Both Wesson and Halevi\footnote{In [17], extending "empty" 5D-GR to (E-non-violating) 
{\bf fluctuations} of ($\pm$) curvatures/densities {\bf around} zero, 
predicts a (hyperlow-probable phase-
transitional) inflation-collapse process, where the 5D-negative densities hyperexplode into 
a 5D-hyperspace (thus integrating a physical, self-diminishing positive cosmological constant 
into the process) and the exploded positive densities collapse into the 4D-
hypersurface/Einstein's spacetime (gradually {\bf and} gracefully).} 
predict our 4DU to be ``a hypersurface in a flat 5D
manifold" [3].  As long as it is {\bf induced} by mathematical equivalences it might be
``an effect of a choice of coordinates in a truncated higher-dimensional geometry" [7].
But, if it was physically produced by the pretransparent dimensional
collapse-these processes do not leave us (confused by) choices: They simultaneously
show how both our hypersurface {\bf and} our XD-blind observations/distorted frame 
of reference are
formed!  Thus, neither ``the big bang is an artifact produced by an unfortunate choice
of coordinates" [7], nor are our photon-dependent 4D-observations a (free) choice.
We are just collectively afflicted by the ``cylindricity-sight disorder"...
Hence -- if ``photonland" is found valid -- it also eliminates unclearities
such as ...the Hubble expansion, the microwave background, and primordial nuclearsynthesis...
are in a sense recognized as geometrical illusions -- artifacts of a choice of coordinates 
in a higher-dimensional world" [5].

The photonland hypothesis must eventually refute or integrate another strange possibility:
 any 5D-neo-KKU is deep enough to enable 
$(^{4}_{3})$ -- if not $ (^{5}_{3})$
 -- orthogonal 
collapse processes into 4 groups of photons (and other real particles), 
each group with zero velocity in {\bf another} dimension.
Analyzing
these primeval complicated  5D-plasma processes might easily refute -- 
or gradually reshape/integrate 
the photonland and Wesson \& Liu's hypotheses.  If photonland survives, the three ``alien" 
photon groups -- instead of illuminating our 4DU-- might illuminate our understanding in 
5D-quantum processes [24].

\vspace{7mm}

Acknowledgments:  I am very grateful to Drs. G. K\"albermann HUJ, and A. Yahalom TAU, 
for numerous discussions.


\end{document}